\documentclass[twocolumn,showpacs,preprintnumbers,amsmath,amssymb]{revtex4}
\newcommand{\beq}{\begin{equation}}
\newcommand{\eeq}{\end{equation}}

\usepackage{graphicx}
\usepackage{dcolumn}
\usepackage{bm}

\begin{document}

\preprint{APS/123-QED}

\title{A novel view of plane wave expansion method in photonic crystals}

\author{Young-Chung Hsue}
 \email{ychsu.ep87g@nctu.edu.tw}
\author{Tzong-Jer Yang}
\email{yangtj@cc.nctu.edu.tw}
 \affiliation{Department of Electrophysics, National Chiao-Tung University, Hsinchu, Taiwan,%
 Republic of China}
\date{\today}

\begin{abstract}
We propose a method derived from the simple plane wave expansion that can easily solve
the interface problem between vacuum and a semi-infinite photonic crystal. The method is
designed to find the complete set of all the eigenfunctions, propagating or evanescent,
of the translation operators $\{\mathbf{T}_{\mathbf{R}}\}$, at a fixed frequency. With
these eigenfunctions and their eigenvalues, the transmitted and reflected waves can be
determined. Two kinds of applications are presented for 2D photonic crystals. The first
is a selection rule for determine the normal direction of the vacuum-photonic crystal
interface to achieve the highest attenuation effect at a gap frequency. The second is to
calculate the transmittance and reflectance for a light incident from vacuum to an
semi-infinite photonic crystal. As an example we recalculate a system studied previously
by K. Sakoda et al. and get results in agreement with theirs.
\end{abstract}

\maketitle

Since Yablonovitch \cite{1,2,3} discovered a periodic dielectric structure that has an
absolute gap in the frequency spectrum for electromagnetic waves, the idea of photonic
crystals have attracted great interest. Many phenomena have been predicted theoretically
and many application possibilities have been explored \cite{2,4,5,6}. Corresponding
studies in the early years most authors paid their attention on the frequency spectrum
gaps, and the most convenient method to calculate the band gaps is the plane wave
expansion method \cite{Sakoda,Gu1}. Recently, various kinds new methods have been
proposed to compute some other relevant physical parameters such as transmittance and
penetration depth \cite{7,8,9,17} for a finite system.

In this paper we also address the transmittance and penetration depth problems, but use a
different method that includes all the information of propagating and evanescent modes
getting from the translation operator. We show that by appropriately modifying the
conventional plane wave expansion method we can enlarge its application region, and which
makes it easy to solve the interface problem between vacuum and a semi-infinite photonic
crystal system.

Our method has several advantages. First, both the ``air rods in dielectric" and ``dielectric rods in air"
problems can be solved, without any restriction on the shapes of the rods and position of cutting plane that
separating the semi-infinite photonic crystal region from the air region, which is impossible for the LKKR
method \cite{12,13,14,15,16}. Second, all information getting from the complete set of the eigenfunctions of the
translation operator are used, including both the propagating and evanescent modes. This makes it easy to
analyze and discuss phenomena using the well established knowledge of solid state physics. Third, the finite
size effects such as the resonance behavior of the transmittance curve caused by the finite thickness of the
photonic crystal sample can be easily isolated. We can thus accurately calculate the transmittance and
reflectance for a very thick photonic crystal sample.

For a system without free charge and current, and if the permittivity $\epsilon({\bf r})$
and permeability $\mu({\bf r})$ are scalars independent of time, the magnetic field ${\bf
H}({\bf r},t)$ satisfies \beq
\nabla\times\frac{1}{\epsilon}\nabla\times\mathbf{H}(\mathbf{r},t)=-\mu\frac{\partial^2}{\partial
t^2}\mathbf{H}(\mathbf{r},t), \eeq where $\mathbf{H}(\mathbf{r},t)=\sum\limits_\omega
\mathbf{H}_\omega(\mathbf{r})e^{-i\omega t}$, and \beq
\nabla\times\frac{1}{\epsilon}\nabla\times\mathbf{H}_{\omega}(\mathbf{r})
=\mu{\omega^2}\mathbf{H}_{\omega}(\mathbf{r}).\label{homega} \eeq In addition, if
$\epsilon({\bf r})$ and $\mu({\bf r})$ are periodic functions, following the derivation
of Bloch theory, Eq. (\ref{homega}) can be changed to
\begin{align}
 - \sum\limits_{{\mathbf{G}}'} {\left( {{\mathbf{k}} + {\mathbf{G}}} \right)
 \times \epsilon _{{\mathbf{G}} - {\mathbf{G}}'}^{ - 1}
 \left( {{\mathbf{k}} + {\mathbf{G}}'} \right) \times
 {\mathbf{H}}_{{\mathbf{k}},{\mathbf{G}}'}} \notag\\
 = \omega ^2 \sum\limits_{{\mathbf{G}}'} {\mu _{{\mathbf{G}} - {\mathbf{G}}'}
   {\mathbf{H}}_{{\mathbf{k}},{\mathbf{G}}'} },
   \label{eq:1}
\end{align}
where ${\mathbf{H}_{\omega}}\left( {\mathbf{r}} \right) = \sum\limits_{\mathbf{G}}
{e^{i\left( {{\mathbf{k}} + {\mathbf{G}}} \right) \cdot {\mathbf{r}}}
{\mathbf{H}}_{{\mathbf{k}},{\mathbf{G}}} }$, $\epsilon \left( {\mathbf{r}} \right) =
\sum\limits_{\mathbf{G}} {e^{i{\mathbf{G}} \cdot {\mathbf{r}}} \epsilon _{\mathbf{G}} }$,
 $\mu \left( {\mathbf{r}} \right) = \sum\limits_{\mathbf{G}} {e^{i{\mathbf{G}}{\mathbf{r}}} \mu _{\mathbf{G}}
}$, and $\{\bf G\}$ is the set of the reciprocal lattice. Since in this paper we consider
only two-dimensional cases, we have $k_z = 0$, and the electromagnetic waves can be
decoupled as E polarization (TE) and H polarization (TM) modes. For example, the TM mode
of ${\bf H}$ field is written as ${\bf H}=H_z\hat{z}$ and satisfy
\begin{align}
\sum\limits_{{\mathbf{G}}'} {\epsilon _{{\mathbf{G}} - {\mathbf{G}}'}^{ - 1} \left(
{{\mathbf{k}} + {\mathbf{G}}} \right) \cdot \left( {{\mathbf{k}} + {\mathbf{G}}'}
\right)H_{z,{\mathbf{k}},{\mathbf{G}}'}}  \notag\\ =  \omega ^2
\sum\limits_{{\mathbf{G}}'} {\mu _{{\mathbf{G}} - {\mathbf{G}}'}
H_{z,{\mathbf{k}},{\mathbf{G}}'} }.
    \label{eq:2}
\end{align}

\begin{figure}[t]
\includegraphics[width=.35\textwidth]{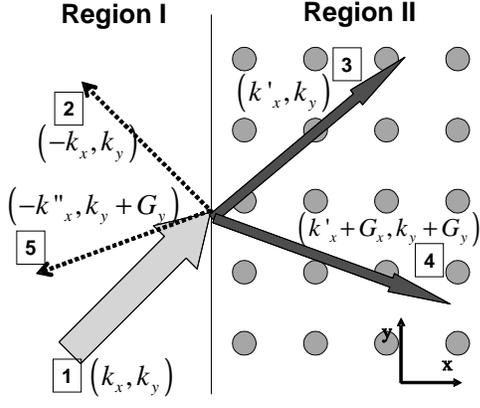}
\caption{\label{fig:1} A schematic view of the light incident from region I to region II,
where region I is vacuum and region II is PC. The gray, dotted, and black arrows
represent the incident, reflected, and transmitted light, respectively. For the incident
light, $\mathbf{k}=(k_x,k_y)$. According to Bloch theory, the modes of region II can be
written as
$H_z(\mathbf{r})=\sum_\mathbf{G}H_\mathbf{G}e^{i(\mathbf{k'+G})\cdot\mathbf{r}}$. Based
on the continuity conditions at the interface, reflection and transmission modes have
$\mathbf{k}_{ref}=(-k"_x,k_y+G_y)$ and $\mathbf{k}_{trans}=(k'_x+G_x,k_y+G_y)$, where
$k''_x=\sqrt{\frac{\omega^2}{c^2}-(k_y+G_y)^2}$, and $k'_x$'s are obtained from solving
Eq.~(\ref{eq:4}).}
\end{figure}

Conventionally, treating Eq.~(\ref{eq:2}) as an eigenvalue problem, the propagating bulk
modes of an infinite periodic system for a given real ${\bf k}$ can be obtained
straightforwardly. However, in most of the situations we also need to know the
transmittance and reflectance of a finite or semi-infinite system for a incident light.
In order to obtain these quantities, various methods such like LKKR \cite{8,9,11},
transfer matrix \cite{12,13,14,15}, and scattering matrix \cite{16} method have been
proposed. In these kinds of methods, a photonic crystal slab is treated as the stack of
many gratings. The matrix problem for only one grating layer is solved first, then
multiplying the matrices layer by layer and the total transmittance and reflectance can
be determined. Although these methods are successful, however, in order to confirm the
good numerical accuracy, the number of layers should be restricted to a small value. In
addition, using these methods it is hard to find the relations between the transmittance
and the original band structure.

In this paper, we propose an alternative method to calculate the transmitted and
reflected waves from the interface between vacuum and a semi-infinite photonic crystal
for a incident plane wave. The method is based on Eq.~(\ref{eq:2}), which contains all
the information of the band structure of the system. We thus can easily analyze the
physical meanings of various phenomena using the knowledge getting from the solid state
physics.

Here, if the $k_x$ and $k_y$ are fixed, the frequency $\omega$ can be solved as an
eigenvalue problem. If the system is infinitely extended, there are just only propagation
modes that can survive. However, sometimes we have to calculate the transmission and
reflection coefficients for a finite sized or semi-infinite sample, which have at least
one boundary. At the edges of the sample, the periodic structure is broken and the
evanescent modes must be considered. However, it is impossible to obtain the evanescent
solutions from the eigenvalue problem of Eq. (\ref{eq:2}) because it provides the
solutions for an extended bulk and the boundary conditions at infinite restrict the $k_x$
and $k_y$ to be taken as real values.

Nevertheless, can we make some modifications to produce the attenuated solutions of
Eq.~(\ref{eq:2})? The answer is yes. For different purposes, there are two kinds of
calculations: one is to fix the direction of {\bf k} and frequency; the other is to fix
$k_y$ and frequency. With a simple transformation, the Eq.~(\ref{eq:2}) can be rewritten
as
\begin{widetext}
\begin{align}
\left( {\begin{array}{*{20}c}
   0 & {{\mathbf{\hat I}}}  \\
   {\epsilon _{{\mathbf{G}} - {\mathbf{G}''}} \left[ {\omega ^2
   \mu _{{\mathbf{G}''} - {\mathbf{G}}'}  - \epsilon _{{\mathbf{G}''}
   - {\mathbf{G}}'}^{ - 1} {\mathbf{G}''} \cdot {\mathbf{G}}'} \right]}
   & { - \epsilon _{{\mathbf{G}} - {\mathbf{G}''}} \left[
   {\epsilon _{{\mathbf{G}''} - {\mathbf{G}}'}^{ - 1} {\mathbf{\hat k}}
   \cdot \left( {{\mathbf{G}''} + {\mathbf{G}}'} \right)} \right]}  \\
 \end{array} } \right)\left( \begin{gathered}
  H_{\mathbf{G}'}  \hfill \\
  kH_{\mathbf{G}'}  \hfill \\
\end{gathered}  \right) = k\left( \begin{gathered}
  H_{\mathbf{G}'}  \hfill \\
  kH_{\mathbf{G}'}  \hfill \\
\end{gathered}  \right)
\label{eq:3}
\end{align}
for fixed direction and frequency, where (respectively) $H_{\mathbf{G'}}$,
$\hat{\mathbf{I}}$ and $\hat{\mathbf{k}}$ denote the abbreviation of
$H_{z,\mathbf{k,G'}}$,$\delta_{\mathbf{G,G'}}$ and unit vector of $\mathbf{k}$, and
\begin{align}
\left( {\begin{array}{*{20}c}
   0 & {{\mathbf{\hat I}}}  \\
   {\epsilon _{{\mathbf{G}} - {\mathbf{G}''}} \left[ {\omega ^2
   \mu _{{\mathbf{G}''} - {\mathbf{G}}'}  - \epsilon _{{\mathbf{G}''}
   - {\mathbf{G}}'}^{ - 1} \left({\mathbf{G}''+k_y\mathbf{\hat y}}\right)
    \cdot \left({\mathbf{G}'+k_y\mathbf{\hat y}}\right)} \right]}
   & \mathbf{\hat P}  \\
 \end{array} } \right)\left( \begin{gathered}
  H_{\mathbf{G}'}  \hfill \\
  k_xH_{\mathbf{G}'}  \hfill \\
\end{gathered}  \right) = k_x\left( \begin{gathered}
  H_{\mathbf{G}'}  \hfill \\
  k_xH_{\mathbf{G}'}  \hfill \\
\end{gathered}  \right)
\label{eq:4}
\end{align}

\end{widetext}
for fixed $k_y$ and frequency, where $\hat{\mathbf{y}}$ and $\mathbf{\hat P}$ denote unit
vector of y direction and ${ - \epsilon _{{\mathbf{G}} - {\mathbf{G}''}} \left[
   {\epsilon _{{\mathbf{G}''} - {\mathbf{G}}'}^{ - 1}\left(
   {{\mathbf{G}''_x} + {\mathbf{G}'_x}}\right) } \right]}$, respectively.
If the ${\bf k}$ and ${\bf G}$ in Eq. (\ref{eq:3}) are acted by a rotation operator
$\hat{\Theta}$ which rotates $\hat{\mathbf{k}}$ to x-direction --- i.e.
$\hat{\Theta}\,\hat{\mathbf{k}}=\hat{\mathbf{x}}$ --- and we define
$\tilde{\mathbf{G}}\equiv\hat{\Theta}\,\hat{\mathbf{G}}$, then Eq. (\ref{eq:3}) becomes
Eq. (\ref{eq:4}) in which $\tilde{\mathbf{G}}$ is substituted for $\mathbf{G}$ and
$k_y=0$. Following this cue, the Eq. (\ref{eq:3}) can be considered as a master equation
for solving a problem where the incident light is always perpendicular to the interface.
From which we can easily determine the penetration depth along direction $\hat{\bf k}$.
If our purpose is to use the band gap effect of the structure, then the result obtained
from this calculation will tell us how to cut the sample to get the highest performance.

\begin{figure}[t]
\includegraphics[width=.35\textwidth]{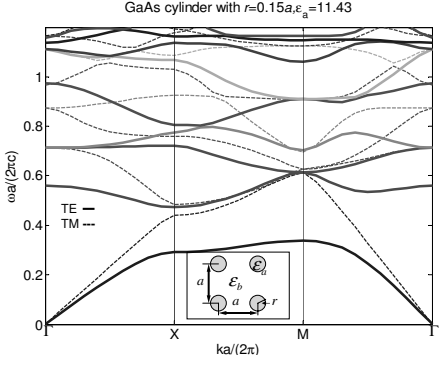}
\caption{\label{fig:2} The frequency spectrum of the square lattice photonic crystal is derived from
Eq.~(\ref{eq:2}). In this figure, solid lines and dash lines denote the TE and TM modes, respectively.}
\end{figure}

On the other hand, Eq.~(\ref{eq:4}) can be used to deal with the problem for light
incident with different angle $\theta$, which is the angle between the normal vector of
the interface and  $\mathbf{k}$ of the incident light. Here
$\tan\theta=k_y/\sqrt{\omega^2/c^2-k_y^2}$. Figure~\ref{fig:1} explains the details.

Since we can obtain all the eigenvectors of the system, the transmission and reflection
spectra can also be obtained. Based on the continuity conditions at the interface, the
relationship between the ${\bf H}$ fields in region I and region II can be written as
\begin{widetext}
\begin{align}
\left( {\begin{array}{*{20}c}
   { - \left\langle {x_0 y|H^{\text{I}}_{m} } \right\rangle } & {\left\langle {x_0 y|H^{{\text{II}}}_{m} } \right\rangle }  \\
   { - \left\langle {x_0 y|\epsilon ^{ - 1} \partial _x |H^{\text{I}} _{m} } \right\rangle } &
   {\left\langle {x_0 y|\epsilon ^{ - 1} \partial _x |H^{{\text{II}}}_{m} } \right\rangle }  \\
 \end{array} } \right)\left( \begin{gathered}
  \left\langle {H^{\text{I}}_{m} |{\mathbf{\hat R}}|H^{\text{I}}_{0} } \right\rangle  \hfill \\
  \left\langle {H^{\text{II}}_{m} |{\mathbf{\hat T}}|H^{\text{I}}_{0} } \right\rangle  \hfill \\
\end{gathered}  \right) = \left( \begin{gathered}
  \left\langle {x_0 y|H^{\text{I}}_{0} } \right\rangle  \hfill \\
  \left\langle {x_0 y|\epsilon ^{ - 1} \partial _x |H^{\text{I}}_{0} } \right\rangle  \hfill \\
\end{gathered}  \right),
\label{eq:5}
\end{align}
\end{widetext}
where $\mathbf{\hat R}$ and $\mathbf{\hat T}$ are the reflection and transmission
operators, $H^\text{I}_{m}$ and $H^\text{II}_{m}$ are the reflection modes in region I
and transmission modes in region II, the $m$ denotes the different modes, respectively.
And $H^\text{I}_{0}$ is the incident light.

If Eq.~(\ref{eq:5}) is expanded in K-space, it can be rewritten as
\begin{align}
\hat{\mathbf{A}}\left( \begin{gathered}
  \left\langle {H_{\text{I} m} |{\mathbf{\hat R}}|H_{\operatorname{I} 0} } \right\rangle  \hfill \\
  \left\langle {H_{{\text{II}}m} |{\mathbf{\hat T}}|H_{\operatorname{I} 0} } \right\rangle  \hfill \\
\end{gathered}  \right) = \left( \begin{gathered}
  H_{z,{\mathbf{k}},{G_y}} ^{{\text{I}},i}   \hfill \\
  k_{G_y,x}^{{\text{I,}}i}H_{z,{\mathbf{k}},{G_y}} ^{{\text{I}},i}   \hfill \\
\end{gathered}
\right), \label{eq:6}
\end{align}
where
$$\hat{\mathbf{A}}= \left( {\begin{array}{*{20}c}
   { - {H_{z,{\mathbf{k}},{G_y}} ^{{\text{I}},r} } } & {\sum\limits_{G_x }
   {H_{z,{\mathbf{k}},{\mathbf{G}}} ^{{\text{II}},t} } }  \\
   { - { k_{mx}^{\text{I}}
   H_{z,{\mathbf{k}},G_y} ^{{\text{I}},r} } } &
   {\sum\limits_{G_x ,{\mathbf{G}}'} {\epsilon _{{\mathbf{G}} - {\mathbf{G}}'}^{ - 1}
    \left( {k_{mx}^{{\text{II}}}  + G_x' } \right)
    H_{z,{\mathbf{k}},{\mathbf{G}}} ^{{\text{II}},t} } }
 \end{array} }
\right), $$
 and $H_{z,\mathbf{k},\mathbf{G}}$ can be gotten from Eq.~(\ref{eq:4}), and
$i,r$ and $t$ denote incident, reflection and transmission, respectively. To determine
the transmission and reflection coefficients, we have to first decide the direction of
the Poynting vector of every mode. For $\mathbf{k}$ is a real vector, it can either be
obtained from $ {\bf v}_g = \vec \nabla _{\mathbf{k}} \omega$ or from
\begin{align}
& \int\limits_{cell} {\operatorname{Re} \left\{ { - \frac{i} {{\omega \epsilon }}H_z^*
\vec \nabla H_z } \right\}} dr^2 \notag\\ &= \sum\limits_{{\mathbf{G}},{\mathbf{G}}'}
{\operatorname{Re} \left\{ {\frac{1} {\omega }H_{z,{\mathbf{k}},{\mathbf{G}}}^* \epsilon
_{{\mathbf{G}} - {\mathbf{G}}'}^{ - 1} \left( {{\mathbf{k}} + {\mathbf{G}}'}
\right)H_{z,{\mathbf{k}},{\mathbf{G}}'} } \right\}} \label{eq:7}
\end{align}

For $\mathbf{k}$ is a complex number, propagation toward right hand side corresponds to
$\operatorname{Im}(k_x)>0$. When group velocity and $H_{z,\mathbf{k},\mathbf{G}}$ of each
mode are known, the transmittance $\mathbf{T}$ and reflectance $\mathbf{R}$ can be gotten
from them, and the accuracy can be estimated from how $\mathbf{R}+\mathbf{T}$ is close to
one. Sometimes, $\det|\mathbf{A}|$ is possible to become zero, if so, the Eq.
(\ref{eq:6}) has nonezero solutions when there is no incident light. This kind of wave is
a surface state which resembles the surface plasmon propagating along the surface of a
metal. However, we do not discuss it in this paper. One more interesting thing among
these three equations --- from Eq.(\ref{eq:2}) to Eq.(\ref{eq:4}) --- is that they are
identical to each other. Because the second row of matrices of left side of Eq.
(\ref{eq:3}) and Eq. (\ref{eq:4}) are equal to Eq. (\ref{eq:2}) while the
$\{\mathbf{G}\}$ in these equations are equal. Thus any eigenfunction of one of these
three equations satisfies another two equations. It leads to two useful things: (i) the
real {\bf k} contours of these three methods with equal frequency are the same. (ii) The
$\{\mathbf{G}\}$ needn't to be changed when these three equations are considered as a
series of policy tools. For example, we should decide where the band edge is and select
the $\omega_0$ near the band edge to obtain the band structure from Eq. (\ref{eq:2}) if
we hope to observe what happened near the band gap. By replacing the frequencies of Eq.
(\ref{eq:3}) and (\ref{eq:4}) by $\omega_0$, the penetration depth can be derived from
Eq. (\ref{eq:3}) and so do the variations of transmittance while the solutions of Eq.
(\ref{eq:4}) are used in Eq. (\ref{eq:5}). During this process, even if we just select
$\mathbf{G}=0$ and drop out $\mathbf{G}\neq0$ for average assumption in the calculations
of Eq.~(\ref{eq:2}), the $\{\mathbf{G}\}$ still needn't to be changed in the following
calculations.

\begin{figure}[t]
\includegraphics[width=.35\textwidth]{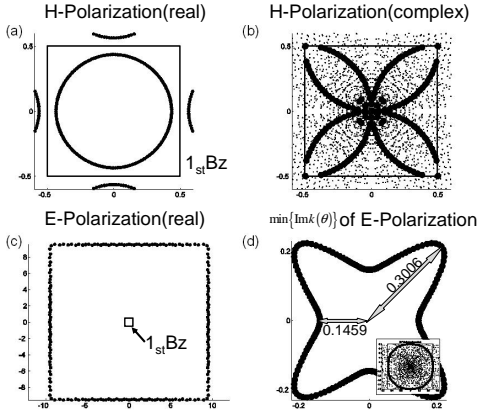}
\caption{\label{fig:3} Possible $\mathbf{k}$ values for a constant-frequency
$\omega=0.4$. The solid square frames in these figures are the first Brillouin zone
boundaries and the data taken are inside the first Brillouin zone. (a) and (b) are for
the TM waves (H-polarization), (c) and (d) are for the TE waves (E-polarization). (a) and
(c) are constant-frequency contours (with purely real ${\bf k}$) of propagating waves.
(b) and the inset of (d) show the contours of Re$\{\mathbf{k}\}$ of evanescent modes. In
addition, (d) shows the $\min\{\operatorname{Im}\tilde{k}(\theta)\}$. The spots in the
(b) and the inset of (d) correspond to the cases with smallest Im$({\bf k})$ or longest
penetration depth.}
\end{figure}

\begin{figure}[t]
\includegraphics[width=.35\textwidth]{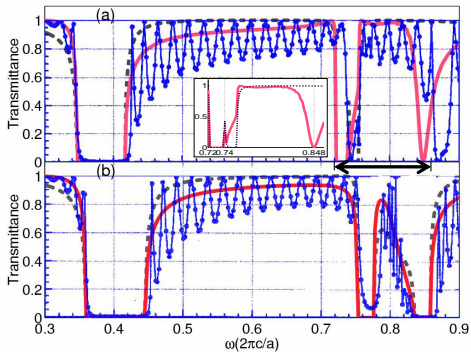}
\caption{\label{fig:4} Transmittance of the (a)TE and (b)TM modes. The line with dots is
the data excerpted from Ref. \cite{7}. The solid lines and dashed lines are our solutions
with different cut planes (in the solid line case the cut plane is the same as in
\cite{7}, whereas in the dashed line case the cut plane passes through the centers of the
cylindrical holes), and the inset is the clearer view of (a) whose frequency regime is
indicated by $\leftrightarrow$.}
\end{figure}

For simplicity, the structure we used in obtaining Fig.~\ref{fig:2} and Fig.~\ref{fig:3}
is the square lattice with GaAs ($\epsilon=11.43$ ) cylinders each with radius=$0.15a$
inside vacuum; whereas in Fig.~\ref{fig:4} we employ vacuum cylinders each with radius
$0.43077a$ inside the PbO($\epsilon=2.72$ ) background. In all cases, $a$ is the lattice
constant and the primitive vectors are given by ${\bf a}_1=(a,0)$ and ${\bf a}_2=(0,a)$.

The first application is a selection rule to determine the interface direction for the
highest performance of light insulation. We solve Eq.~(\ref{eq:3}) at a frequency equal
to $0.4 (2\pi c/a)$ and the results are presented in Fig.~\ref{fig:3}. In
Fig.~\ref{fig:2}, there is a band gap at $\omega=0.4(2\pi c/a)$ for the TE mode, so that
in Fig.~\ref{fig:3}(c), there are no real number solutions inside the first Brillouin
zone. However, outside and far away from the 1st Brillouin zone such solutions exist,
which are fake and are caused by the finite basis used in the calculations. To find the
interface direction we first choose a direction $\hat{\bf k}$ and use Eq.~(\ref{eq:3}) to
find a $k$ that has the smallest $|{\rm Im}(k)|$ value, which denoted as $k_I(\theta)$
and determines the main decay trend for a wave propagating along $\hat{\bf k}$. The
second step is to scan angles from $0^\circ$ to $45^\circ$ to find an angle $\theta_0$
that has the maximum $k_I(\theta)$. The details are shown in Fig.~\ref{fig:3}(d), where
we calculate the TE modes, and the penetration depths (i.e., $2\pi/k_I(\theta)$) for
$0^\circ$ and $45^\circ$ are $6.8540 a$ and $3.3272 a$, respectively. This indicates that
when we produce a sample that cut along $45^\circ$, it just needs 4 or 5 layers to stop
the light with $\omega=0.4(2\pi c/a)$ for TE modes instead of 7 or 8 layers for
$0^\circ$.

The second application is to fix $k_y$ and frequency in order to obtain $k_x$. For
comparison, we select the system discussed in \cite{7} to contrast with our system and
show the result in Fig.~\ref{fig:4}. The system in \cite{7} is a 16 layers photonic
crystal which is a square lattice (with a lattice constant $a$) of air columns (radius
equals $0.43077a$ and is located at the center of a unit cell) in a dielectric substrate
PbO placed in air and our system is a semi-infinite photonic crystal placed in air. By
using Eq.~(\ref{eq:4}) and Eq.~(\ref{eq:6}) the transmittance can be obtained as shown in
Fig.~\ref{fig:4}. Because our system is infinite, we can find something quite different.

(1) The solid lines and dashed lines are almost smooth curves except in the gap regime and at some special
points ($\omega= 0.74$ and $0.85$ in TE mode). The oscillating solid lines with dots in Fig. 4 represent the
solutions of \cite{7}. It is obvious that our curves are different from that of \cite{7}. The oscillation
behavior is owing to finite thickness. They can be easily explained by a roughly consideration of the average
dielectric $\bar{\epsilon}\equiv \left<\epsilon\right>_{cell}$. For low frequency, the most important
contribution of $\epsilon_\mathbf{G}$ is $\epsilon_{\mathbf{G}=0}$, which is equal to $\bar{\epsilon}$,
therefore the $16$ layers photonic crystals can be considered as an effective material with uniform dielectric
$\bar{\epsilon}=1.7173\epsilon_0$ and width $16a$, where $a$ is the lattice constant of the photonic crystal and
$\epsilon_0$ is the permittivity of free space. This is a typical 1D problem and the waves in both sides of this
material can be connected by transfer matrix whose formula is $$\mathbf{M}_\omega \left(
\begin{array}{cc} e^{ik16a} & 0 \\ 0 & e^{-ik16a} \\
\end{array}\right)\mathbf{M}_\omega^{-1},$$ where $\mathbf{M}_\omega$ is a $2\times 2$
matrix with $\det|\mathbf{M}_\omega|\neq 0$ determined by the optical impedance contrast
and the incident angle, and $k=\omega\sqrt{\bar{\epsilon}\mu_0}$ is the effective wave
number, where $\mu_0$ is permeability of free space. It is obvious that the transfer
matrices equals $\pm\mathbf{I}$ when $k16a=n\pi$, where $n$ is an integer number. That
means the frequency difference between two neighboring peaks is
$$\Delta\tilde{\omega}=\frac{1}{32\sqrt{\bar{\epsilon}/\epsilon_0}}=
\frac{1}{32\sqrt{1.7173}}\simeq 0.0238, $$ where $\tilde{\omega}=\frac{\omega a}{2\pi
c}$. Comparing with the average width $0.0222$ of peaks between $\frac{\omega a}{2\pi
c}=0.5$ to $0.7$, we can say the oscillation in Fig. 4 comes from the effect of finite
size.

(2) In the vicinity of $\tilde{\omega}\simeq 0.74$, according to the band structure
calculation results the spectrum should ascend rapidly when $\tilde{\omega}$ is
increasing, because it is at band edge. But, the real situation appears in Fig. 4(a) is
ascending quickly and descending immediately to near zero transmittance. Our explanation
is that there are two propagation mode's with group velocity $v_g\simeq 0$, so they do
not contribute to the transmittance.

(3) In Fig.~\ref{fig:4}(a), the valley of transmittance near $\omega=0.85$ disappears
when the interface is chosen to pass through the center of the vacuum cylinders. This
reveals that it is possible to stop the light at some isolated frequency points by
appropriately choosing the cutting plane of the photonic crystal even if the frequencies are
outside of band gaps.

(4) In Fig.~\ref{fig:4}(b), there should be a forbidden band for $\omega=0.75$ to 0.78,
but the line with dots does not show this result. According our calculation the wave
attenuation rate for a 16-layer structure is about 0.0733, which agrees with the result
shown by the line with dots. This phenomenon shows that the evanescent modes do
contribute the transmittance in a finite thickness structure.

In these two applications, the evanescent modes are necessary and useful for the
calculations of semi-infinite system, and this method also provides us some information
such like for how large a separation distance between two defects can they be treated as
independent in super-cell method.

In conclusion, the method we present here may not be efficient enough, because in the
calculation we get results both inside and outside of the first Brillouin zone (FBZ).
However, only the results inside the FBZ are useful and the others repeat the same
information and are redundant. For example, if we use $N^2$ bases, there are only $2N$
useful eigenfunctions. Besides, this method has several advantages. First, from this
method we can easily realize and analyze some properties of periodic systems with
interface and the computational time is independent of the number of layers. Thus, even
if the number of layers is very large, it will save much time. Second, using
Eq.~(\ref{eq:7}), we can also calculate the density of states, D($\omega$), through
\begin{align}
D\left( \omega  \right) = \int\limits_{shell} {\frac{{dk_{//} }} {{\left| {\nabla
_{\mathbf{k}} \omega } \right|}}}. \label{eq:8}
\end{align}
They are especially useful when we aim to calculate the density of states in some small
frequency regimes.

We are now investigating the cases of finite size specimens and a structure with line
defects.

  Finally, we thank Prof. B. Y. Gu for instructing us about Andreev reflection, which gave us a chance to
  employ this idea about Eq.~(\ref{eq:3}), and thank Dr. P. G. Luan who let us find more
  possibilities with this method.

\end{document}